\documentclass[twocolumn,aps,prb,showpacs]{revtex4}

\usepackage{epsf}
\usepackage{bm}

\begin{document}

\title{
Size--sensitive melting characteristics of gallium clusters:
Comparison of Experiment and Theory for Ga$_{17}{}^{+}$ and
Ga$_{20}{}^{+}$
}

\author{Sailaja Krishnamurty, S. Chacko, and D. G. Kanhere}

\affiliation{
Department of Physics, and
Centre for Modeling and Simulation,
University of Pune,
Ganeshkhind,
Pune--411 007,
India}

\author{G. A. Breaux, C. M. Neal, and M. F. Jarrold}

\affiliation{
Chemistry Department,
Indiana University,
800 East Kirkwood Avenue,
Bloomington,
Indiana 47405--7102}

\date{\today}

\begin{abstract}

Experiments and simulations have been performed to examine the
finite-temperature behavior of Ga$_{17}{}^{+}$ and Ga$_{20}{}^{+}$
clusters.
Specific heats and average collision cross sections have been measured
as a function of temperature, and the results compared to simulations
performed using first principles Density--Functional
Molecular--Dynamics.
The experimental results show that while Ga$_{17}{}^{+}$ apparently
undergoes a solid--liquid transition without a significant peak in the
specific--heat, Ga$_{20}{}^{+}$ melts with a relatively sharp peak.
Our analysis of the computational results indicate a strong
correlation between the ground--state geometry and the
finite--temperature behavior of the cluster.
If the ground--state geometry is symmetric and ``ordered" the cluster
is found to have a distinct peak in the specific--heat.
However, if the ground--state geometry is amorphous or ``disordered"
the cluster melts without a peak in the specific--heat.

\end{abstract}


\maketitle

\section{Introduction}
\label{sec.intro}

It is now well established that the melting points of particles with
thousands of atoms decrease smoothly with decreasing particle size, due
to the increase in the surface to volume ratio.~\cite{Pawlow, Nature-1977}
However, unlike particles with thousands of atoms or the bulk material,
probing the finite--temperature properties of small clusters (with
$<$ 500 atoms) is non--trivial and remains a challenging task.
Experimental studies of the melting transitions of clusters in small
size regime have only recently become possible.~\cite{Haberland-1997,
Haberland-Nature, Haberland-PRL-2003, Haberland-PRL-2005, Jarold-Tin,
Jarrold-Gallium1, Jarrold-Jacs, Jarrold-Al, Jarrold-Tinfrag}
Several interesting phenomena have been observed, including
melting temperatures that rise above the bulk
value~\cite{Jarold-Tin,Jarrold-Gallium1} and strong size--dependent
variations in the melting temperatures.~\cite{Jarrold-Jacs, Jarrold-Al}.
These experimental findings have motivated many theoretical
investigations on the finite--temperature behavior of
clusters.~\cite{Ga-prl, Our-PRBsn10, Our-PRBsn20, James-Tin,
Eur-Phys.J, Na-PRB, Na-JCP}
Simulations based on first principles have been particularly
successful in quantitatively explaining the factors behind the
size--dependent variations in the melting behavior of
clusters.~\cite{Na-PRB}
Thus, a confluence of recent advances in experimental methods and
theoretical studies using first principles methods have set the stage
for a major increase in our understanding of phase transitions in
these small systems.

Coming to the present work on gallium clusters, it is by now well
known that gallium clusters not only melt at substantially higher
temperatures than the bulk ($T_{m[\rm bulk]}=$
303~K),~\cite{Jarrold-Gallium1} but they also exhibit wide variations
in the temperature dependences of their specific--heats, with some
clusters showing strong peaks (due to the latent heat), while others
(apparent "non-melters") showing no peak.~\cite{Jarrold-Jacs}.
These features show a strong dependence on cluster size, where the
addition of a single atom can change a cluster with no peak in
the specific--heat into a ``magic melter" with a very distinct peak.
This behavior has been observed for gallium clusters, Ga$_{n}$,
with $n=$ 30--55.

In the present work, we probe the melting behavior of small gallium
cluster ions where we show that the ``non-melting" and ``melting"
features in the specific--heats are observed in clusters as small
as Ga$_{17}{}^{+}$ and Ga$_{20}{}^{+}$, respectively.
Prior experimental results for Ga$_{17}^{+}$ over a limited temperature
range showed no evidence for a melting
transition.~\cite{Jarrold-Gallium1}
The experimental results in this case were specific--heat measurements
performed using multi--collision induced dissociation, where a peak in
the specific--heat due to the latent heat was the signature of melting.
On the other hand, recent simulations for Ga$_{17}$ show a broad peak
in the specific--heat centered around 600~K.
The previous specific--heat measurements for Ga$_{17}^{+}$ extended
only up to 700~K, so one possible explanation for this apparent
discrepancy is that the melting transition occurred at a slightly
higher temperature than examined in the experiments.
Here, we report specific--heat measurements for Ga$_{17}^{+}$ over a
more extended temperature range, along with specific--heat
measurements for Ga$_{20}^{+}$.
While no peak is observed in the heat--capacities for Ga$_{17}^{+}$, a
peak is observed for Ga$_{20}^{+}$.

To further probe the melting transitions in these clusters, ion
mobility measurements were performed for Ga$_{17}^{+}$ and
Ga$_{20}^{+}$ as a function of temperature.
The ion mobility measurements provide average collision cross sections
which can reveal information about the
shape and volume changes that occur on melting.
For example, a cluster with a non--spherical geometry might be
expected to adopt a spherical shape (a liquid droplet) on melting.
If there is not a significant shape change, there may still be a
volume change on melting.
Most bulk materials expand when they melt (the liquid is less dense
than the solid).
Even in the absence of a significant shape or volume change, the cross
sections might show an inflection at the melting transition due to the
thermal coefficient of expansion of the liquid cluster being larger
than for the solid (in the macroscopic regime most liquids have larger
coefficients of expansion than the corresponding solids).
An inflection is observed in the cross sections for both Ga$_{17}^{+}$
and Ga$_{20}^{+}$.
Thus, the ion mobility measurements suggest that Ga$_{17}^{+}$ as
well as Ga$_{20}^{+}$ are in a liquidlike state above 800~K.

To explore the reasons behind the behavior outlined above (i.e., to
determine why Ga$_{17}^{+}$ apparently melts without a peak in its
specific--heat, while a peak is observed for
Ga$_{20}^{+}$), we have carried out first principles
Density--Functional~(DF) Molecular--Dynamics (MD) calculations on both
clusters.
The ground--state structure and the bonding within the
clusters is analyzed.
The ionic specific--heat is computed using multiple histogram
method.~\cite{MH,amv-review}
The calculated specific--heats for Ga$_{17}^{+}$ show three broad
low intensity maxima that extend from 300 to 1400 K.
This resembles the experiment results where
the measured specific--heats are relatively featureless.
In contrast, the calculated specific--heats for Ga$_{20}^{+}$ show
a clear peak around 750~K.
This is in excellent agreement with the peak obtained from
experimental measurements (which occurs at around 700~K).
Finally, our theoretical results show that the features in the
specific--heat curves are influenced by the ground--state geometry,
the bonding of the atoms within the ground--state structure,
and the isomer distribution that becomes accessible as the
temperature is raised.

In Sec.~\ref{sec:method}, we present the experimental methods and the
computational details.
In Sec.~\ref{sec:rd} we discuss the experimental and theoretical
results on both clusters.
We conclude our results in Sec.~\ref{sec:concl}.

\section{Methodology\label{sec:method}}

Specific-heats were measured using the recently developed
multi--collision induced dissociation approach.
The cluster ions are generated by laser vaporization of a liquid
gallium target in a continuous flow of helium buffer gas.
After exiting the laser vaporization region of the source the
clusters travel through a 10 cm long temperature variable
extension where their temperature is set. Cluster ions that
exit the extension are focused into a quadrupole mass spectrometer
where a particular cluster size is selected.
The size selected clusters are then focused into a collision cell
containing 1~Torr of helium.
As the clusters enter the collision cell they undergo numerous
collisions with the helium, each one converting a small fraction of
the ions translational energy into internal energy.
If the initial kinetic energy is high enough some of the cluster ions
may be heated to the point where they dissociate.
The dissociated and undissociated cluster ions are swept across the
collision cell by a small electric field and some of them exit through
a small aperture.
The ions that exit are analyzed in a second quadrupole mass
spectrometer and then detected by an off--axis collision dynode and
dual microchannel plates.
The fraction of the ions that dissociate is determined from the mass
spectrum.
Measurements are performed as a function of the ions initial kinetic
energy, and the initial kinetic energy required for 50\% dissociation
(IKE50\%D) is determined from a linear regression.
IKE50\%D is measured as a function of the temperature of the
temperature--variable extension on the source.
IKE50\%D decreases as the temperature is raised because hotter
clusters have more internal energy, and hence less energy needs to be
added in order to cause
dissociation.
At the melting transition a sharp decrease in IKE50\%D is expected due
to the latent heat.
The derivative of IKE50\%D with respect to temperature is approximately
proportional to the specific--heat.
The proportionality constant is the fraction of the clusters initial
kinetic energy that is converted into internal energy, which is
estimated from an impulsive collision model.
A drop in the IKE50\%D values due to the latent heat of a melting
transition leads to a peak in the specific--heat.

Ion mobility measurements can provide information on the shape and
volume changes that occur when clusters melt.
For the ion mobility measurements, the collision cell is replaced by a
7.6 cm long drift tube.
50 $\mu$s pulses of cluster ions are injected into the drift tube and
the drift time distribution is obtained by recording the ions arrival
times at the detector with a multichannel scalar.
Average collision cross sections are obtained from the drift time
distributions using standard methods.~\cite{masonmcdaniel}

All the simulations are performed using
Born--Oppenheimer molecular--dynamics based on Kohn--Sham formulation
of Density--Functional Theory~(DFT).~\cite{KS}
We have used Vanderbilts' ultrasoft
pseudopotentials~\cite{uspp-vanderbilt} within the GGA approximation,
as implemented in the \textsc{vasp} package~\cite{vasp} for both
clusters.
For all calculations, we use only $4s^2$ and $4p^1$--electrons as
valence, taking the 3$d$--electrons~\cite{d-electron} as a part of the
ionic core.
An energy cutoff of about $\approx~10$~Ry is used for the plane--wave
expansion of the wavefunction, with a convergence in the total energy
of the order of 0.0001~eV.
Cubic supercells of lengths 20 and 25~{\AA} are used for Ga$_{17}^+$
and Ga$_{20}^+$, respectively.
For examining the finite--temperature behavior, the ionic phase space
of the clusters is sampled by isokinetic MD where kinetic energy is
held constant via a velocity scaling method.
For both the clusters, we split the total temperature range from
100--1400~K into 15 different temperatures.
We maintain the cluster at each temperature for a period of at
least 90~ps, leading to total simulation times of the order of 1~ns.
The resulting trajectory data were used to compute standard
thermodynamic indicators as well as the ionic specific--heat, via a
multihistogram technique.
Details can be found in Ref.~\onlinecite{amv-review,abhijat-mh}.

\section{Results and Discussion\label{sec:rd}}

\begin{figure}
  \epsfxsize=0.50\textwidth
  \centerline{\epsfbox{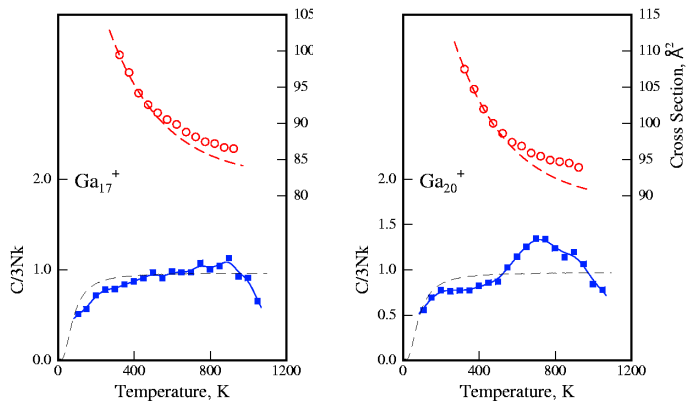}}
  \caption{\label{fig.expt}
 Specific--heats and average collision cross sections
 measured for size--selected Ga$_{17}{}^{+}$ and Ga$_{20}{}^{+}$
 clusters as a function of temperature. The solid blue points show
 the specific--heats which are normalized to $3Nk_B$ (the classical
 value), where $k_B$ is the Boltzmann constant and $N = (3n-6+3/2)/3$
 ($n$ = number of atoms in the cluster, and $3n-6$ and 3/2 are due
 to the vibrational and rotational contributions, respectively).
  }
\end{figure}

Specific--heats measured for Ga$_{17}^+$ and Ga$_{20}^+$ as a
function of temperature are shown in the lower half of
Fig.~\ref{fig.expt}.
The points are the experimental values, while the
dashed line is the prediction of a modified Debye model.
In the case of Ga$_{17}^+$, the specific--heats shown in
Fig.~\ref{fig.expt} appear to gradually increase up to around 900~K.
The sharp decrease in the specific--heats above 900~K is an artifact
due to evaporative cooling, the spontaneous unimolecular dissociation
of the cluster ions as they travel between the source extension and
the collision cell.
For Ga$_{20}^+$, the specific--heats show a broad maximum, around
400~K wide, centered at around 725~K.
The peak for Ga$_{20}^+$ is significantly broader than observed for
larger clusters (like Ga$_{39}^+$ and Ga$_{40}^+$) where the peak was
attributed to a melting transition.
However, it is well known that the melting transition, and the
corresponding peak in the specific--heats, becomes broader with
decreasing cluster size.
Thus, even though the peak in the specific-heats for Ga$_{20}^+$ is
around 400~K wide, it is appropriate to assign it to a finite--size
analog of a bulk melting transition.
The center of the peak is at around 725~K, this is well above the bulk
melting point (303~K).
This continues a trend reported for larger cluster sizes ($n=$ 30--55)
where the melting temperatures are also significantly above the bulk
value.
The unfilled red circles in Fig.~\ref{fig.expt} show the average
collision cross sections determined for Ga$_{17}^+$ and Ga$_{20}^+$ as
a function of the temperature.
The cross sections are expected to systematically decrease with
increasing temperature because the long range attractive interactions
between the cluster ion and the buffer gas atoms becomes less
important, and the collisions become harder as the temperature is
raised.
The thick dashed red line in the figures show the expected exponential
decrease in the cross sections with increasing temperature.
There is an inflection in the cross sections for Ga$_{20}^+$ that
appears to slightly precede the peak in the specific--heat for this
cluster.
The inflection in consistent with a melting transition where the
liquid cluster has a larger coefficient of thermal expansion than the
solid.
There is also an inflection in the cross sections for Ga$_{17}^+$.
This suggests that a solid--liquid transition also occurs for
Ga$_{17}^+$, but without a significant peak in the specific--heat.

\begin{figure}
  \epsfxsize=0.4\textwidth
  \centerline{\epsfbox{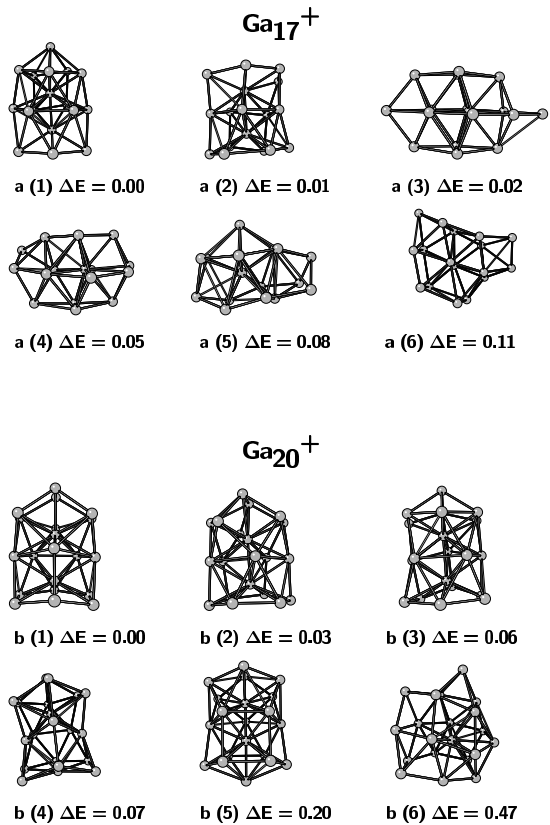}}
  \caption{\label{fig.geom}
  The ground--state and some representative low--lying and
  excited--state geometries of Ga$_{17}{}^{+}$ and Ga$_{20}{}^{+}$
  clusters.
  Fig (1) corresponds to the ground--state geometry.
  The energy difference $\Delta E$ is given in eV with respect to the
  ground--state.
  }
\end{figure}

To understand the reason behind the different behavior observed for
Ga$_{17}{}^{+}$ and Ga$_{20}{}^{+}$, we have carried out a detailed
analysis of structure and bonding in both clusters.
As will be become apparent from the following discussion, that
the ground--state geometry and the nature of bonding plays a crucial
role in determining the finite--temperature behavior of the cluster.
We begin with a discussion of the ground--state geometries of
cationic Ga$_{17}$ and Ga$_{20}$ clusters.
We have obtained more than 50 distinct equilibrium geometries by
quenching more than 200 structures, selected from a few high
temperature MD runs, for both sizes.
In Fig.~\ref{fig.geom}, we show the lowest energy structure along with
some low lying excited state geometries of both clusters.
The lowest energy geometry of the Ga$_{17}{}^{+}$ cluster~(see
Fig.~\ref{fig.geom}--a(1)) is similar to that of Ga$_{17}$ reported in
our earlier work.~\cite{Ga-prl}
It has a distorted decahedral structure, which suggests the
possibility of further cluster growth to a 19--atom double decahedron.
In contrast, the ground--state geometry of Ga$_{20}{}^+$, shown
in Fig.~\ref{fig.geom}--b(1), is more symmetric.
It can be described as a double decahedral structure of 19 atoms,
with the bottom capped atom merging into the pentagonal plane to form
a hexagonal ring.
In addition, an atom from the top pentagon and the upper capped atom
rearrange to accommodate the 20$^{\rm th}$ atom, leading to a
dome--shaped hexagonal ring.

We now analyze the structural properties in detail to get an insight
into the features that influence the melting characteristics.
An analysis of the bond--length distribution shows that there are
12 bonds, for each cluster, having distances less than
2.55~{\AA}.~\cite{metallic-bond}
Interestingly, for Ga$_{17}{}^+$, these short bonds are spread all
over the cluster, whereas for Ga$_{20}{}^+$, they form the upper and
the lower hexagonal rings.
The distribution of coordination numbers~\cite{coordination-number}
indicate that for Ga$_{20}{}^+$, almost all the atoms in the rings
(about 16), have a coordination number of 4.
The Ga$_{17}{}^+$ cluster, however, does not have such a uniform
distribution of coordination numbers.
Thus, the ground--state geometry of Ga$_{17}{}^+$ might be considered to
be ``disordered", while that of Ga$_{20}{}^+$ exhibits a more--ordered
structure.

\begin{figure}
  \epsfxsize=0.4\textwidth
  \centerline{\epsfbox{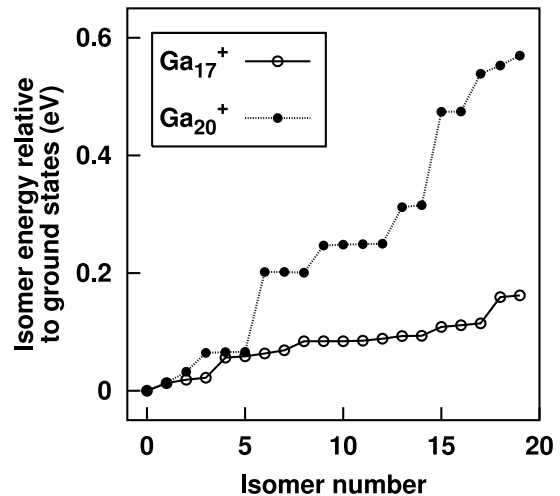}}
  \caption{\label{fig.isomer}
  The energies of the isomeric structures of Ga$_{17}{}^{+}$ and
  Ga$_{20}{}^{+}$ with respect to their ground--states.
  }
\end{figure}

Striking differences are also observed in the low
energy isomers and their distribution on the potential--energy surface.
As mentioned above, we have obtained more than 50 distinct isomers
spanning an energy range of about 1.0~eV above the ground--states for
each cluster.
In Fig.~\ref{fig.isomer}, we plot the energies of these isomers
relative to the ground--state, arranged in an ascending order.
The isomers for the Ga$_{17}{}^+$ cluster appear to exhibit an
almost continuous energy distribution.
While a few of these isomers are severe distortions of the
ground--state geometry, the rest do not show any resemblance~(see
Fig.~\ref{fig.geom}a).
It appears that for the Ga$_{17}{}^+$ isomers in this low energy
regime, small rearrangements of the atoms, costing just a small
amount of energy, lead to several close-lying isomers, so that
the isomer distribution is almost continuous.
In contrast, the isomers of Ga$_{20}{}^+$ cluster are distributed
in three groups, separated by an energy gap of about
0.2~eV~(Fig.~\ref{fig.isomer}).
The first group of isomers have slightly different orientations of
atoms in the hexagonal rings and are nearly degenerate with the
ground--state.
The second group consists of structures having only the lower
hexagonal ring while the third group has no rings.
This indicates that the hexagonal units of Ga$_{20}{}^+$ cluster are
stable and difficult to break.
The stability of the ring--pattern of Ga$_{20}{}^+$ and the isomer
distribution for both clusters should have a substantial effect on
the melting characteristics.
Indeed, as we shall see further below, these features play a
crucial role in the finite--temperature characteristics.
It should be mentioned that although these observations are based on
rather limited search, we believe that the general features described
here are essentially correct.

\begin{table}
\caption{
\label{table.elf}
The number of basins with more than one atom at different values
of the
electron localization function for the ground--state structures
of Ga$_{17}{}^+$ and Ga$_{20}{}^+$ clusters.
The numbers in parenthesis represent the number of atoms in each
basin.
}

\begin{tabular}{ccc}
\hline
                                 ELF value
\hspace{0.0cm} & \hspace{1.0cm}  Ga$_{17}{}^+$
\hspace{1.0cm} & \hspace{1.0cm}  Ga$_{20}{}^+$        \\
\hline
    0.85      &  0              &    1 (2)            \\
    0.77      &  1 (2)          &    2 (5,7)          \\
    0.75      &  3 (2,2,2)      &    2 (5,7)          \\
    0.73      &  2 (3,4)        &    1 (14)           \\
\hline
\end{tabular}
\end{table}

The most important difference between the two clusters is the nature
of the bonding.
We use the concept of an electron localization
function~(ELF),~\cite{elf-silvi} to describe the nature of bonding.
This function is normalized to a value between zero and unity; a value
of 1 represents a perfect localization of the valence charge while the
value for the uniform electron gas is 1/2.
The locations of maxima of this function are called \emph{attractors},
since other points in space can be connected to them by paths of
maximum gradient.
The set of all such points in space that are attracted by a maximum is
defined to be the \emph{basin} of that attractor.
Basin formations are usually observed as the value of the ELF is
lowered from its maximum, at which there are as many basins as the
number of atoms in the system.
Typically, the existence of an isosurface or a basin in the bonding
region between two atoms at a high ELF value, say $\ge 0.70$,
signifies a localized bond in that region.

\begin{figure}
  \epsfxsize=0.4\textwidth
  \centerline{\epsfbox{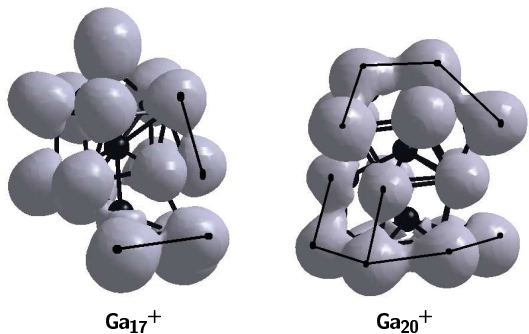}}
  \caption{\label{fig.elf}
  The isosurface for the electron localization functions for
  Ga$_{17}{}^+$   and Ga$_{20}{}^+$ at an isovalue of 0.75.
  The black lines correspond to merged basin structures.
  }
\end{figure}

We have analyzed the electron localization functions for Ga$_{17}{}^+$
and Ga$_{20}{}^+$ clusters for values $\le0.85$.
In Table~\ref{table.elf}, we give the number of basins containing
two or more atoms, for selected ELF values.
The table clearly shows a fragmented growth pattern of the basins for
Ga$_{17}{}^+$, each containing very few atoms as compared to that of
Ga$_{20}{}^+$.
For instance, at an isovalue of 0.75, while Ga$_{17}{}^+$ has three
basins each having 2 atoms, Ga$_{20}{}^+$ has just two basins each
containing 5 and 7 atoms that corresponds to the two hexagonal rings.
The ELF contours for the isovalue of 0.75 are shown in
Fig.~\ref{fig.elf}.
The merged basins structures are shown by the black lines.
It may be inferred that the bonds between atoms in the hexagonal
rings of Ga$_{20}{}^+$ are strong and covalent in nature with similar
strengths, while the fragmented basin growth pattern in Ga$_{17}{}^+$
indicates inhomogeneity of the bond strengths.

\begin{figure}
  \epsfxsize=0.4\textwidth
  \centerline{\epsfbox{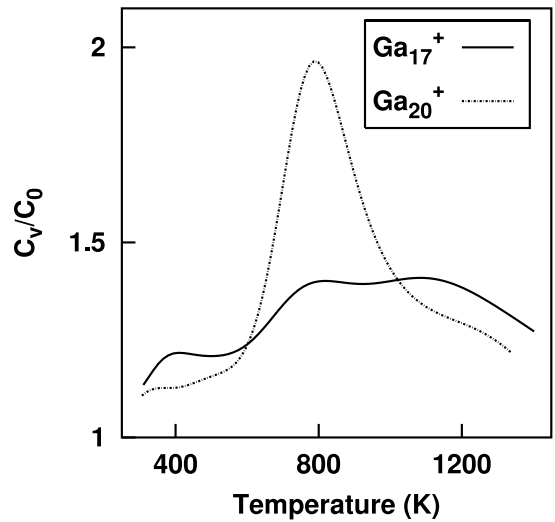}}
  \caption{\label{fig.cv}
  Normalized canonical specific--heat for
  Ga$_{17}{}^+$ (continuous--line) and Ga$_{20}{}^{+}$ (dashed--line).
  $C_0=(3n-6 + 3/2)k_B$ is the zero temperature classical limit of the
  rotational plus vibrational canonical specific--heat.
  }
\end{figure}

The calculated, normalized, canonical specific--heats
are shown in Fig.~\ref{fig.cv} plotted against temperature.
The plot for Ga$_{17}{}^+$ exhibits a broad feature (apparently
consisting of three components) which extends from 300~K to 1400~K.
For Ga$_{20}{}^+$, the calculated specific--heat remains nearly
flat up to about 600~K, it then increases
sharply and peaks at about 800~K, in excellent agreement with
the experimental results described above.
Thus, interesting size--sensitive features seen in the experimental
heat--capacities are reproduced in our simulations.
This behavior can be understood from our earlier discussion of the
bond--length distributions, coordination numbers,
isomer--distributions, and the nature of bonding in these clusters.
While the Ga$_{17}{}^+$ cluster shows no real evidence for ordered
behavior, the
Ga$_{20}{}^+$ cluster has well--ordered ring--patterns.
Thus, when Ga$_{17}{}^+$ is heated, the bonds soften
gradually, and the cluster hops through all its isomers continuously.
This is clearly demonstrated by the ionic motion as a function of
temperature, which shows that this cluster evolves through all
isomers smoothly from 300~K to 1400~K.
On the other hand, the ionic motion for Ga$_{20}{}^+$ shows only
minor rearrangements of the atoms until 600~K, and then the cluster
visits all the isomers corresponding to the first group of isomers
described above.
At about 700~K, the upper hexagonal ring breaks, while at about 800~K,
the lower ring breaks.
Thus, melting of Ga$_{20}{}^+$ cluster is associated with the breaking
of the well--ordered covalently bonded hexagonal units.

\begin{figure}
  \epsfxsize=0.4\textwidth
  \centerline{\epsfbox{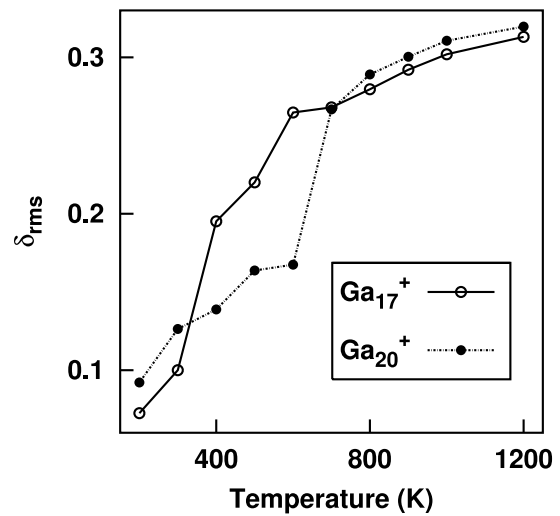}}
  \caption{\label{fig.delta}
  Root--mean--square bond--length--fluctuations ($\delta_{\rm rms}$)
  for Ga$_{17}{}^+$ (continuous--line) and
  Ga$_{20}$ (dashed--line).
  }
\end{figure}

We have also analyzed the melting characteristics via traditional
parameters such as, the root--mean--squared
bond--length--fluctuations~($\delta_{\rm rms}$) and the mean--squared
ionic displacements (MSD).
In Fig.~\ref{fig.delta}, we show the $\delta_{\rm rms}$ for
Ga$_{17}{}^+$ and Ga$_{20}{}^+$ clusters.
This plot correlates well with the specific--heat curve shown in
Fig.~\ref{fig.cv}.
The $\delta_{\rm rms}$ for Ga$_{17}{}^+$ rises gradually from 300~K,
while for Ga$_{20}{}^+$, it rises sharply at about 700~K, and finally
saturates to the same value for both clusters.
It may be inferred from this observation that the behavior of both
clusters at temperatures say, $T \ge 800$~K, are similar and that
both clusters can be considered to be in \emph{liquidlike} states.
This conclusion is further substantiated by the MSD plots (figures not
shown), which saturate at $\approx$ 21$\AA^2$, at about 1200~K for both
clusters.

\section{Summary and conclusion\label{sec:concl}}

It is evident from the present study that the
nature of the ground--state geometry and bonding strongly influences
the finite--temperature characteristics of Ga$_{17}{}^+$ and
Ga$_{20}{}^+$.
At high temperatures, $T \ge 800$~K, both
Ga$_{17}{}^+$ and Ga$_{20}{}^+$ have similar root--mean--squared
bond--length--fluctuations and the mean--squared ionic displacements
so that both of them can be considered to be in liquidlike states.
The experimental results show that while Ga$_{17}{}^{+}$ apparently undergoes
a solid--liquid transition without a significant peak in
the specific--heat, Ga$_{20}{}^{+}$ melts with a relatively sharp peak.
The simulations show that if the cluster is ``ordered" (i.e. a large
fraction of the constituent
atoms show similar bonding, coordination numbers, and bond energies)
then it is likely to show a sharp melting transition with a
significant peak in the specific--heat. On the other hand,
if the cluster is ``disordered" (i.e. the constituent atoms occur in
a wide distribution of bonding environments) it will probably undergo
a solid--liquid transition without a significant peak in the
specific--heat. In the latter case, the number of isomers or
conformations sampled by the cluster increases steadily as the
temperature is raised, instead of the abrupt increase that occurs when
a cluster undergoes a sharp melting transition.

These observations have interesting consequences for the
finite--temperature behavior of small clusters as a function of
cluster size.
It is likely that as clusters grow in size their structures evolve
from one well--ordered structure to another, passing on the way
through some cluster sizes that have ``disordered" structures.
For instance, the 13--atom gallium cluster is predicted to have a
highly symmetric decahedron structure with a bonding pattern that is
similar to that found here for Ga$_{20}{}^+$.~\cite{Ga-prl}
So in the present case, cluster growth from Ga$_{13}$ (a decahedron)
to (Ga$_{20}{}^+$, a distorted double--decahedron) proceeds via a
disordered Ga$_{17}{}^+$ structure.
Such behavior is also observed for sodium clusters in 40 to 55 atom
size range; the ground--state geometries of Na$_{40}$ and Na$_{55}$
are either icosahedron or close to icosahedron while that of Na$_{50}$
has no particular symmetry.~\cite{Na-PRB,Na-JCP}
In such cases, we expect that the specific--heats should change from
showing a well--defined peak to a rather broad one, and back again to
well--defined.
We believe this behavior to be generic as it has not only been
observed in case of gallium
clusters~\cite{Jarrold-Gallium1,Jarrold-Jacs} but also in case of
aluminum clusters~\cite{Jarrold-Al}, experimentally, and for sodium
clusters in the simulations mentioned above.

\section{Acknowledgment}

The financial support of the Indo--French Center for
Promotion of Advanced Research is greatfully acknowledged.
C--DAC (Pune) is acknowledged for
providing us with the supercomputing facilities.
We gratefully acknowledge partial support of this work by the US National
Science Foundation.


\begin{thebibliography}{}
\bibitem{d-electron}
 In our earlier work on gallium clusters~[see
 Ref.~\onlinecite{Ga-prl}], we have verified that the $d$--electrons
 do not significantly affect the equilibrium geometries and the
 finite--temperature behavior.
\bibitem{metallic-bond}
 A typical bond--distance of a metallic bond seen in the extended
 system, $i.e.$ $\alpha$--gallium, is about 2.79~{\AA}.
\end{thebibliography}
\end{document}